\documentclass{JHEP3} 
\usepackage{latexsym} 
\usepackage{amsmath}

\title{New Results on Holographic Three-Point Functions}

\author{Massimo Bianchi and Maurizio Prisco\\ 
        Dipartimento di Fisica, \ Universit\`a di Roma ``Tor
        Vergata'' and \\
        I.N.F.N.\ -- Sezione di Roma ``Tor Vergata'', \\
        Via della Ricerca Scientifica 1, 00133  Roma,  Italy\\
        E-mail: \email{massimo.bianchi@roma2.infn.it},
        \email{maurizio.prisco@roma2.infn.it}} 

\author{Wolfgang M\"uck \\
        Dipartimento di Scienze Fisiche,
	Universit\`a di Napoli ``Federico II'' and \\
	I.N.F.N.\ -- Sezione di Napoli, \\
	Via Cintia, 80126 Napoli, Italy \\ 
	E-mail: \email{mueck@na.infn.it}}

\preprint{ROM2F/03/27 \\NA-DSF-35-2003 \\hep-th/0310129}

\keywords{AdS-CFT Correspondence, Renormalization Group}

\abstract{We exploit a gauge invariant approach for the analysis of
  the equations governing the dynamics of active scalar fluctuations
  coupled to the fluctuations of the metric along holographic RG
  flows. In the present approach, a second order ODE for the active
  scalar emerges rather simply and makes it possible to use the
  Green's function method to deal with (quadratic) interaction
  terms. We thus fill a gap for active scalar operators, whose
  three-point functions have been inaccessible so far, and derive a
  general, explicitly Bose symmetric formula thereof. As an
  application we compute the relevant three-point function along the
  GPPZ flow and extract the irreducible trilinear couplings of the
  corresponding superglueballs by amputating the external legs
  on-shell.}

\begin{document}

 
\providecommand{\ie}{\emph{i.e.\ }} 
\providecommand{\eg}{\emph{e.g.\ }} 
 
\providecommand{\rmd}{\mathrm{d}}  
\providecommand{\dx}{\rmd x} 
\providecommand{\dr}{\rmd r} 
\providecommand{\e}[1]{\mathrm{e}^{#1}} 
 
\providecommand{\F}{\mathrm{F}} 
 
\providecommand{\LeP}{\mathrm{P}}

\providecommand{\K}{\mathcal{K}} 
 
\providecommand{\tg}{\tilde{g}} 
\providecommand{\tR}{\tilde{R}} 
 
\providecommand{\tG}[2]{\tilde{\Gamma}^{#1}\!_{#2}} 
\providecommand{\G}[2]{\Gamma^{#1}\!_{#2}} 
 
\providecommand{\bg}{\bar{g}} 
\providecommand{\bp}{\bar{\phi}} 
 
\providecommand{\hh}{\hat{h}} 
\providecommand{\rh}{\check{h}} 
\providecommand{\hp}{\hat{\varphi}} 
\providecommand{\rp}{\check{\varphi}} 
\providecommand{\ha}{\hat{a}} 
\providecommand{\ra}{\check{a}} 
 
\providecommand{\Gone}[2]{\Gamma^{(1)#1}{}_{#2}} 
 
\newcommand{\Ofn}[1]{\mathcal{O}(f^{#1})} 
\newcommand{\Of}{\mathcal{O}(f)} 
 
\providecommand{\htt}{{h^{TT}}} 
 
\providecommand{\tE}{\tilde{E}} 
 
\providecommand{\ta}{\tilde{a}} 
 
\providecommand{\tK}{\tilde{K}} 

\providecommand{\X}{\mathcal{X}}
\providecommand{\Y}{\mathcal{Y}}
 
\providecommand{\vev}[1]{\left\langle #1 \right\rangle} 


\section{Introduction} 
\label{intro} 
The holographic calculation of correlation functions in conformal 
field theories has been pursued much in recent years, spawned by  
the formulation of the AdS/CFT correspondence 
\cite{Gubser:1998bc,Witten:1998qj}. Roughly speaking, the on-shell 
action of a (super)gravity theory living in a $(d+1)$-dimensional 
bulk 
space-time can be interpreted as the generating functional of a dual 
quantum field theory (QFT), if the bulk geometry is asymptotically 
anti-de Sitter (AdS). The prescribed boundary values of the bulk 
fields play 
the role of the QFT sources. If the bulk geometry is
entirely AdS, the dual QFT is conformal. Otherwise, the bulk 
geometries
describe renormalization group (RG) flows of the QFT away from the
ultraviolet conformal fixed point, which are driven by relevant
operators, either by deforming the action or by turning on non-zero 
vacuum
expectation values (vevs). These are called operator and vev RG flows,
respectively. The correspondence formula has been given a precise 
meaning by the development of a systematic renormalization procedure,
known as holographic renormalization, that removes all infrared 
divergences
stemming from the infinite volume in the bulk on-shell action and
ensures the validity of the (anomalous) Ward identities in the
boundary QFT \cite{deHaro:2000xn, Henningson:1998gx, Bianchi:2001kw,
  Martelli:2002sp, Skenderis:2002wp}. 
 
The most famous duality starts with gauged supergravity in 
$(d+1)=5$ dimensions, which is obtained from $D=10$ type IIB 
supergravity 
by compactification on a five-sphere. Its $AdS_5$ solution is dual to 
$d=4$, 
$\mathcal{N}=4$ super Yang-Mills (SYM) theory in the planar 
limit at strong 't Hooft coupling \cite{Maldacena:1998re}. Several RG
flows of this theory, with various degrees of supersymmetry, have 
been identified \cite{Khavaev:1998fb, Girardello:1998pd, 
Freedman:1999gp, 
  Behrndt:1999ay, Brandhuber:1999jr, Freedman:1999gk, Behrndt:1999kz, 
  Girardello:1999bd, Evans:2000ap, Petrini:1999qa}.  
 
Explicit calculations of $n$-point functions in RG flows with $n>2$ 
have been hampered by a number of reasons. First, in RG flow 
backgrounds, the fluctuations of the active scalar\footnote{In the 
  common nomenclature, an \emph{active} scalar is dual to the 
  operator driving the RG flow and has a non-zero background, whereas the other
  scalars are called \emph{inert}.} 
couple to the fluctuations of the bulk metric in a  
non-trivial fashion even at the linearized level, and much effort was 
needed in previous work to disentangle this mixing and to 
extract some two-point functions 
\cite{DeWolfe:2000xi, Arutyunov:2000rq, Bianchi:2000sm, Muck:2001cy,  
Bianchi:2001de}. Second, for many cases the background is such that 
the equations of motion are not explicitly solvable  
even for the simplest fields, \ie inert scalars and transverse modes  
of the graviton and vector bosons. The 
favourite solvable cases are the GPPZ flow \cite{Girardello:1999bd},  
which captures some features of the flow from $\mathcal{N}=4$ to 
$\mathcal{N}=1$ 
super Yang-Mills (SYM) theory, and the Coulomb branch flow
\cite{Freedman:1999gk}, which breaks $SU(4)$ R-symmetry   
to $SO(4)$. 
The difficulties make approaching the calculation of higher-point 
functions rather daunting. However, interesting mass spectra have 
been exposed by two-point functions, and the explicit knowledge of 
three-point vertices would be very desirable. The first calculation 
of 
some very simple three-point functions of operators dual to inert 
scalars has been 
presented in \cite{Bianchi:2003bd}.   
 
In this paper, we consider the active scalar more thoroughly with the 
aim of providing a 
general and simple formula for the three-point function of its dual 
operator. Thanks to a Ward identity, this operator is equivalent to 
the trace of the QFT energy-momentum tensor in the case of operator 
RG 
flows. We shall apply the result to the GPPZ flow, which is an 
operator flow driven by the insertion of an operator of UV dimension 
$\Delta=3$.  
 
Tackling Einstein's equations to quadratic order around the RG flow 
background is quite an enterprise. Choosing a gauge to remove 
some of the fields is not of much help, since one cannot simply guess 
which gauge leads to the simplest equations for the remaining 
fields. Therefore, we develop and use a gauge invariant method. The 
idea is the following. From the various fields one can form a number 
of independent gauge invariant combinations (call them collectively
$I$), which represent the true degrees of freedom. Using 
diffeomorphism 
invariance, Einstein's equations can then be equivalently expressed 
in 
terms of $I$ only. This means that in the expansion to second order 
those fields that represent gauge artifacts can be dropped. Although 
this seems a lot like choosing a gauge, the equations to be solved 
are, in fact, gauge independent. Much to our surprise, this procedure 
leads to a second order ODE for the active scalar without much 
effort. Hence, interactions can be dealt with in the standard fashion 
by using the Green's function, and this makes three-point 
functions in principle accessible. It is, of course, impossible in 
most cases to perform the final bulk integral involving three 
bulk-to-boundary propagators. Nevertheless one can extract 
irreducible vertices by  
amputating on-shell external legs.

To finish the introduction, let us briefly introduce the problem and
outline the rest of the paper. 
Our task will be to analyze the dynamics of bulk gravity coupled to a 
scalar field.  
Einstein's equations can be conveniently cast into the
form\footnote{For our notation, see Sec.~\ref{geom}}
\begin{equation} 
\label{Einstein} 
  E_{\mu\nu} = \tR_{\mu\nu} + 2 S_{\mu\nu} = 0~, 
\end{equation} 
where  
\begin{equation} 
  S_{\mu\nu} = \partial_\mu \phi \partial_\nu \phi +\frac{2}{d-1} 
  \tg_{\mu\nu} V(\phi)~. 
\end{equation} 
Bianchi's identity implies the equation of motion for the scalar, 
\begin{equation} 
\label{scalar_eq} 
  \tilde{\nabla}^2 \phi -\frac{\partial V}{\partial \phi} =0~. 
\end{equation} 
 
We analyze the equations of motion using the time slicing formalism, 
in which the bulk metric takes the form  
\begin{equation} 
\label{tslice_metric} 
  \rmd s^2 = (n^2+n_in^i) \dr^2 + 2 n_i \rmd r \rmd x^i  
             + g_{ij} \rmd x^i \rmd x^j~. 
\end{equation} 
A brief summary of the time slicing formalism is presented in 
Sec.~\ref{geom}, which can be skipped in a cursory reading.  
The time slicing formalism has the virtue that the equations are 
expressed covariantly in terms of hyper surface quantities, which 
makes the calculations less prone to errors.  
 
We wish to consider fluctuations around a supergravity background that
is dual to an RG flow. Therefore, we expand the fields as follows, 
\begin{equation} 
\label{expand} 
\begin{split} 
  \phi   &= \bp(r) +\varphi~,\\ 
  g_{ij} &= \e{2A(r)} \left( \eta_{ij} + h_{ij} \right)~,\\ 
  n_i    &= \nu_i~,\\ 
  n      &= 1+ \nu~, 
\end{split} 
\end{equation} 
where $\varphi$, $h_{ij}$, $\nu_i$ an $\nu$ denote the small 
fluctuations. The background functions $\bp(r)$ and $A(r)$ satisfy 
gradient flow equations
governed by a superpotential $W(\phi)$ according to \cite{Freedman:1999gp} 
\begin{equation} 
\label{background} 
\begin{split} 
   \partial_r \bp(r) &= W_\phi~,\\ 
   \partial_r A(r)   &= -\frac{2}{d-1} W~,\\ 
    \frac12 W_\phi^2 - \frac{d}{d-1} W^2 &=V~. 
\end{split} 
\end{equation} 
Here and henceforth, we denote $W_\phi=\rmd W(\phi) / \rmd \phi 
|_{\bp}$, and similar for higher derivatives. 
Furthermore, we adopt the convention that the indices of the 
fluctuations as well as of the derivatives $\partial_i$ 
are raised and lowered with the flat (Minkowski/Euclidean) metric.

In Sec.~\ref{gauge_inv}, we develop the gauge invariant approach that 
starts by identifying gauge invariant combinations, $I$, of the 
fluctuations $\varphi$, $\nu$, $\nu^i$ and $h^i_j$. It is then 
demonstrated that Einstein's equations can be equivalently written in 
terms of $I$ only, and a simple recipe is given how to achieve this 
goal.  
The gauge invariant field equations governing the dynamics of the 
active scalar are given explicitly in Sec.~\ref{field_eq}. Two of the 
three coupled equations are easily solved, and the remaining equation 
for the active scalar is a linear second order ODE with higher order 
source terms.   
In Sec.~\ref{corr}, we derive a general formula for the three-point 
function of the operator that is dual to the active scalar. It is 
given in terms of an integral over the radial bulk variable involving 
three 
bulk-to-boundary propagators and an operator involving the external 
momenta and radial derivatives. Some effort is needed to 
arrive at an expression that is explicitly Bose symmetric. The 
formula 
is applied to the GPPZ flow, in which case one can perform the 
integral, if all external legs are on-shell and amputated.  
Finally, Sec.~\ref{conc} contains our conclusions. Useful details for 
the GPPZ flow are listed in appendix~\ref{GPPZ_rels}. 
 
\section{Geometric Relations for Hyper Surfaces} 
\label{geom} 
The time slicing (or ADM) formalism \cite{Wald,MTW}, which we will 
employ in our analysis of Einstein's equations, makes essential use 
of 
the geometry of hyper surfaces \cite{Eisenhart}. Therefore, we shall  
begin with a review of the basic relations governing the 
geometry of hyper surfaces.

A hyper surface in a 
space-time with coordinates $X^\mu$ ($\mu=0,\ldots, d$) and metric 
$\tg_{\mu\nu}$ is defined by a set of $d+1$ functions, $X^\mu(x^i)$ 
($i=1,\ldots, d$), where the $x^i$ are a set of (local) coordinates 
on the 
hyper surface. The tangents, $X^\mu_i \equiv \partial_i 
X^\mu$, and the normal vector, $N^\mu$, to the hyper surface satisfy 
the 
following orthogonality relations, 
\begin{equation} 
\label{geom:ortho} 
\begin{split} 
  \tg_{\mu\nu}\, X^\mu_i X^\nu_j &= g_{ij}~,\\ 
  \tg_{\mu\nu} X^\mu_i N^\nu &= 0~,\\ 
  \tg_{\mu\nu} N^\mu N^\nu &=1~, 
\end{split} 
\end{equation} 
where $g_{ij}$ represents the (induced) metric on the hyper surface.  
Henceforth, a tilde will be used to label quantities 
characterizing the ($d+1$)-dimensional space-time manifold, whereas 
those of the hyper surface remain unadorned.  
 
The equations of Gauss and Weingarten define the second fundamental 
form, $\K_{ij}$, of the hyper surface, 
\begin{align} 
\label{geom:gauss1} 
  \partial_i X^\mu_j + \tG{\mu}{\lambda\nu} X^\lambda_i X^\nu_j 
  - \G{k}{ij} X^\mu_k &= \K_{ij} N^\mu~,\\ 
\label{geom:weingarten} 
  \partial_i N^\mu + \tG{\mu}{\lambda\nu} X^\lambda_i N^\nu  
  &= -\K^j_i X^\mu_j~. 
\end{align} 
The second fundamental form describes the extrinsic curvature of the 
hyper surface and is related to the intrinsic curvature by another 
equation of Gauss,\footnote{Our convention for the Riemann tensor is 
$R^i{}_{jkl} = \partial_l \G{i}{jk} + \G{i}{lm}\G{m}{jk} - (k 
    \leftrightarrow l)$.} 
\begin{align} 
\label{geom:gauss2} 
  \tR_{\mu\nu\lambda\rho} X^\mu_i X^\nu_j X^\lambda_k X^\rho_l  
  &= R_{ijkl} - \K_{il} \K_{jk} + \K_{ik} \K_{jl}~.\\ 
\intertext{Furthermore, it satisfies the equation of Codazzi,} 
\label{geom:codazzi} 
  \tR_{\mu\nu\lambda\rho} X^\mu_i X^\nu_j N^\lambda X^\rho_k  
  &= \nabla_j \K_{ik} - \nabla_i \K_{jk}~.  
\end{align} 
The symbol $\nabla$ denotes covariant derivatives with respect to the 
induced metric, $g_{ij}$.   
 
The above formulae simplify, if (as in the familiar time slicing 
formalism), we choose space-time coordinates such that  
\begin{equation} 
\label{geom:Xdef} 
  X^0 = \mathit{const}~, \quad X^i=x^i~. 
\end{equation} 
Then, the tangent vectors are given by $X^0_i=0$ and $X^j_i 
=\delta_i^j$. One conveniently splits up the space-time 
metric as (shown here for Euclidean signature) 
\begin{align} 
\label{geom:split} 
  \tg_{\mu\nu} &= \begin{pmatrix} n_i n^i +n^2 & n_j \\ 
				n_i & g_{ij} \end{pmatrix}~,\\ 
\intertext{whose inverse is given by} 
\label{geom:splitinv} 
  \tg^{\mu\nu} &= \frac1{n^2} \begin{pmatrix} 1& -n^j \\ 
			-n^i & n^2 g^{ij} +n^i n^j \end{pmatrix}~. 
\end{align} 
The matrix $g^{ij}$ is the inverse of $g_{ij}$ and is used to raise 
hyper surface indices. The quantities $n$ and 
$n^i$ are the lapse function and shift vector, respectively.  
 
The normal vector $N^\mu$ satisfying the orthogonality relations  
\eqref{geom:ortho} is given by  
\begin{equation} 
\label{geom:normal} 
  N_\mu = (n,0)~, \qquad N^\mu= \frac1{n}(1,-n^i)~. 
\end{equation} 
Then, one can obtain the second fundamental form from the equation of 
Gauss \eqref{geom:gauss1} as  
\begin{equation} 
\label{sli:Kij} 
  \K_{ij} = - \frac1{2n} \left(\partial_0 g_{ij} - 
  \nabla_i n_j - \nabla_j n_i \right)~. 
\end{equation} 
 
We are interested in expressing all bulk quantities in terms of hyper 
surface quantities. Using the equations of Gauss and Weingarten, some 
Christoffel symbols can be expressed as follows, 
\begin{align} 
\label{geom:conn_0ij} 
  \tG{0}{ij} &= \frac1n \K_{ij}~,\\ 
\label{geom:conn_kij} 
  \tG{k}{ij} &= \G{k}{ij} - \frac{n^k}n \K_{ij}~,\\ 
\label{geom:conn_0i0} 
  \tG{0}{i0} &= \frac1n \partial_i n + \frac{n^j}n \K_{ij}~,\\ 
\label{geom:conn_ki0} 
  \tG{k}{i0} &= \nabla_i n^k - \frac{n^k}n \partial_i n  
  - n\K_{ij} \left( g^{jk}+ \frac{n^jn^k}{n^2} \right)~. 
\end{align} 
The remaining components, $\tG{0}{00}$ and $\tG{k}{00}$, 
are easily found from their definitions using \eqref{geom:split} 
and \eqref{geom:splitinv}, 
\begin{align} 
\label{geom:conn_000} 
  \tG{0}{00} &= \frac1n \left( \partial_0 n +n^j \partial_j n +n^i 
n^j 
  \K_{ij} \right)~,\\ 
\label{geom:conn:k00} 
  \tG{k}{00} &= \partial_0 n^k + n^i \nabla_i n^k - n \nabla^k n -2n 
  \K^k_i n^i  
  -\frac{n^k}{n} \left( \partial_0 n +n^j \partial_j n +n^i n^j 
\K_{ij} 
  \right)~. 
\end{align}

\section{Gauge Invariant Approach} 
\label{gauge_inv} 
 
\subsection{Gauge Transformations and Invariants} 
Invariance under diffeomorphisms is a powerful tool that allows one 
to 
reduce the number of fields in the field equations to the effective 
degrees of freedom, which is usually done by fixing a gauge. In our 
treatment of the coupled scalar-gravity system we shall take a 
slightly different approach identifying gauge invariant quantities, 
in 
terms of which the field equations are expressed. Thereby we obtain 
the equations of motion in a gauge invariant form. 
Alternatively, our treatment indicates a choice of gauge, 
in which the equations of motion become particularly simple.  
 
Let us start with the transformations under a  
diffeomorphism of the form  
\begin{equation} 
\label{diffeo} 
  x^\mu\to x'{}^\mu=x^\mu- \xi^\mu(x)~, 
\end{equation} 
where $\xi$ is infinitesimal.  
Because general covariance implies that Einstein's equations are 
invariant 
under any diffeomorphism, they are invariant under \eqref{diffeo} to 
any order in $\xi$, and it is sufficient to consider the first order 
in $\xi$ only. Under \eqref{diffeo}, a scalar field transforms as 
\begin{equation} 
\label{phitrafo} 
 \delta \phi       = \xi^\mu \partial_\mu \phi~, 
\end{equation} 
whereas a covariant tensor of rank two (\eg the metric or 
Einstein's equations) transforms as  
\begin{equation} 
\label{Etrafo} 
 \delta E_{\mu\nu} = \partial_\mu \xi^\lambda E_{\lambda\nu}   
                    +\partial_\nu \xi^\lambda E_{\mu\lambda}  
		    +\xi^\lambda \partial_\lambda E_{\mu\nu}~. 
\end{equation} 
 
Splitting the fields into background and fluctuations as in 
\eqref{expand}, the transformations \eqref{phitrafo} and 
\eqref{Etrafo} become gauge transformations for the fluctuations, 
\begin{equation} 
\label{trafos} 
\begin{split} 
 \delta \varphi &= W_\phi \xi^r +\Of,\\ 
 \delta \nu &= \partial_r \xi^r +\Of~,\\ 
 \delta \nu^i &= \partial^i \xi^r + \e{2A} \partial_r \xi^i+\Of~,\\ 
 \delta h^i_j &= \partial_j \xi^i +\partial^i (\eta_{jk} \xi^k)  
 -\frac{4}{d-1} W \delta^i_j \xi^r+\Of~. 
\end{split} 
\end{equation} 
By $\Ofn{n}$ we denote terms of order $n$ in the fluctuations 
$\varphi$, $h_{ij}$, $\nu_i$ and $\nu$. 
(Remember that we are always at first 
order in the gauge parameter $\xi$.)  
Moreover, we split the metric fluctuations $h^i_j$ as follows, 
\begin{equation} 
\label{hsplit} 
 h^i_j = \htt^i_j  
  + \partial^i \epsilon_j +\partial_j \epsilon^i 
  + \frac{\partial^i \partial_j}{\Box} H + \frac1{d-1} \delta^i_j h~, 
\end{equation} 
where $\htt^i_j$ denotes the traceless transversal part, and 
$\epsilon^i$ 
is a transversal vector ($\partial_i \epsilon^i=0$). It is 
straightforward to obtain from \eqref{trafos}  
the transformation laws for these components, 
\begin{equation} 
\label{h_trafos} 
\begin{split} 
  \delta \htt^i_j &=\Of~,\\ 
  \delta \epsilon^i &= \Pi^i_j \xi^j+\Of~,\\  
  \delta H &= 2 \partial_i \xi^i+\Of~,\\ 
  \delta h &= -4 W \xi^r+\Of~. 
\end{split} 
\end{equation} 
Here, $\Pi^i_j$ is the transversal projector, 
\begin{equation} 
\label{Pproj} 
  \Pi^i_j = \delta^i_j - \frac{\partial^i \partial_j}{\Box}~.
\end{equation} 
  
Using the transformation laws \eqref{trafos} and \eqref{h_trafos}  
we can construct the following gauge 
invariant combinations of the fluctuations, 
\begin{align} 
\label{A} 
 a &= \varphi +W_\phi \frac{h}{4W}+\Ofn{2}~,\\ 
\label{B} 
 b &= \nu + \partial_r \left( \frac{h}{4W} \right)+\Ofn{2}~,\\ 
\label{C} 
 c &= \partial_i \nu^i + \Box \frac{h}{4W}  
 -\frac12 \e{2A} \partial_r H+\Ofn{2}~,\\  
\label{D} 
 d^i &= \Pi^i_j \nu^j - \e{2A} \partial_r \epsilon^i+\Ofn{2}~,\\ 
\label{E}  
  e^i_j &=\htt^i_j+\Ofn{2}~. 
\end{align} 
Here, $c$ and $d^i$ both stem from $\delta \nu^i$, which we have 
split 
into its longitudinal and transversal parts.  
It is in principle possible to find the higher order terms in 
\eqref{A}--\eqref{E} explicitly, but we shall argue in the next 
subsection that this is not necessary for our purposes. 

We would like to finish this 
subsection by making two crucial observations.  
For convenience, let us arrange the fluctuations into two sets, 
$X=(h,H,\epsilon^i)$, and $Y=(\varphi,\nu,\nu^i,\htt^i_j)$. 
Furthermore, let us collect also the  
invariants to $I=(a,b,c,d^i,e^i_j)$. We shall henceforth use $X$, 
$Y$, 
and $I$ to denote any of the fields of the corresponding 
set. The first observation is that we can write the 
gauge parameter $\xi$ to first order as a linear functional of the 
variations 
of the fields $X$,  
\begin{equation} 
\label{xi_delX} 
  \xi^\lambda = z^\lambda(\delta X) +\Ofn{2} = \delta z^\lambda(X) 
  +\Ofn{2}~. 
\end{equation} 
This is clear from \eqref{h_trafos}.  
Second, the gauge invariant combinations have been chosen such that  
the fields $Y$ can be written in the form  
\begin{equation} 
\label{Yeq} 
  Y = I + y(X) +\Ofn{2}~, 
\end{equation} 
where $y$ is a linear functional of the fields $X$. Moreover, when 
going to the next order in the fluctuations, one can choose $I$ such 
that the quadratic terms do not contain terms with two $I$s, \ie 
\begin{equation} 
\label{Yeq2} 
  Y = I + y(X) + \alpha(X,X) +\beta(X,I) +\Ofn{3}~, 
\end{equation} 
where $\alpha$ and $\beta$ are bilinear in their arguments.  
Let us now turn our attention to the field equations. 
 
\subsection{Einstein's Equations and Gauge Invariance} 
It is our goal to re-write Einstein's equations to quadratic order 
in the fluctuations in terms of the gauge invariant combinations 
$I$. To do this, we start by expanding them symbolically in the form 
\begin{equation} 
\label{E_expand} 
  E_{\mu\nu} = E^{(1)1}_{\mu\nu}(X) + E^{(1)2}_{\mu\nu}(Y) 
              + E^{(2)1}_{\mu\nu}(X,X) + E^{(2)2}_{\mu\nu}(X,Y)  
	      + E^{(2)3}_{\mu\nu}(Y,Y) +\Ofn{3}~. 
\end{equation} 
Here, $E^{(1)}$ and $E^{(2)}$ denote linear and bilinear terms, 
respectively. The background equations are satisfied identically.  
Now we substitute \eqref{Yeq2} for $Y$, which yields 
\begin{equation} 
\label{E_2} 
  E_{\mu\nu} = \tE^{(1)1}_{\mu\nu}(X) + E^{(1)2}_{\mu\nu}(I) 
              + \tE^{(2)1}_{\mu\nu}(X,X) + \tE^{(2)2}_{\mu\nu}(X,I)  
	      + E^{(2)3}_{\mu\nu}(I,I) +\Ofn{3}~. 
\end{equation} 
Notice that the functionals $E^{(1)2}$ and $E^{(2)3}$ are essentially 
unchanged, whereas the others gets modified. For 
example, 
$\tE^{(2)2}$ receives contributions from $E^{(2)2}$, $E^{(2)3}$ and 
$E^{(1)2}$ [through $\beta$ in \eqref{Yeq2}].  
 
Eqn.\ \eqref{E_2} can be vastly simplified by considering its 
transformation under diffeomorphisms.  
From \eqref{Etrafo} and \eqref{xi_delX} we find   
\begin{equation} 
\label{E_trafoX} 
  \delta E_{\mu\nu} = \partial_\mu \delta z^\lambda(X) 
E_{\lambda\nu}   
                    +\partial_\nu \delta z^\lambda(X) E_{\mu\lambda}  
		    +\delta z^\lambda(X) \partial_\lambda E_{\mu\nu} 
                    +\Ofn{3}~. 
\end{equation} 
As there is no first order term on the right hand side, the first 
order terms of $E_{\mu\nu}$ must be invariant, which implies that 
$\tE^{(1)1}_{\mu\nu}(X)=0$ in \eqref{E_2}. We have explicitly checked 
that this is the case.   
Then, substituting $E_{\mu\nu}=E^{(1)2}_{\mu\nu}(I)+\Ofn{2}$  
into the right hand side of \eqref{E_trafoX} yields 
\begin{equation} 
\label{E_trafo2} 
  \delta E_{\mu\nu} = \delta \left[  
                    \partial_\mu z^\lambda(X) 
E^{(1)2}_{\lambda\nu}(I) 
		   +\partial_\nu z^\lambda(X) E^{(1)2}_{\mu\lambda}(I)  
		   +z^\lambda(X) \partial_\lambda E^{(1)2}_{\mu\nu}(I) 
                    \right] +\Ofn{3}~. 
\end{equation} 
Comparing \eqref{E_trafo2} with \eqref{E_2}, we find that  
\begin{equation} 
\begin{split} 
 \tE^{(2)1}_{\mu\nu}&=0~,\\ 
 \tE^{(2)2}_{\mu\nu} &=    
                  \partial_\mu z^\lambda(X) E^{(1)2}_{\lambda\nu}(I) 
		 +\partial_\nu z^\lambda(X) E^{(1)2}_{\mu\lambda}(I)  
                 +z^\lambda(X) \partial_\lambda 
E^{(1)2}_{\mu\nu}(I)~. 
\end{split} 
\end{equation} 
Moreover, every single term in $\tE^{(2)2}_{\mu\nu}$ contains the 
first order equations of 
motion. Therefore, it can be dropped. (We can 
freely use the first order  
equations to simplify the interaction terms). 
Thus, we arrive at the following equation to be solved,  
\begin{equation} 
\label{eqn} 
  E^{(1)2}_{\mu\nu}(I) + E^{(2)3}_{\mu\nu}(I,I) =0~, 
\end{equation}   
which is obtained by the following recipe. \emph{We expand the 
equations of 
motion to second order in the fluctuations $X$ and $Y$. Then, we 
replace every $Y$ by its corresponding $I$ and simply drop all 
$X$s.}  
This simple rule is summarized by the following substitutions, 
\begin{equation} 
\label{field_subs} 
  \phi     \to a~,~ 
  \nu      \to b~,~ 
  \nu^i    \to d^i + \frac{\partial^i c}{\Box}~,
  h^i_j \to e^i_j~. 
\end{equation} 
 
For the sake of completeness, we list here the expressions that 
follow 
from the rules \eqref{field_subs} for the quantities that appear in 
the  
field equations. The extrinsic curvature becomes 
\begin{align} 
\label{nH_exp} 
\begin{split} 
  n\K^i_j &\to \frac{2}{d-1} W \delta^i_j -\frac12 \partial_r e^i_j 
  +\frac12 \e{-2A} \left( \partial^i d_j +\partial_j d^i  
  + 2 \frac{\partial^i \partial_j}{\Box} c \right)  
  +\frac12 e^i_k \partial_r e^k_j \\ 
  &\quad  
  - \frac12 \e{-2A} \left[  
  e^i_k \left( \partial^k d_j +\partial_j d^k + 
  2 \frac{\partial_j \partial^k}{\Box} c \right)  
  + \left( d^k +\frac{\partial^k c}{\Box}  \right) \left( \partial^i 
e_{jk} 
  + \partial_j e^i_k -\partial_k e^i_j \right)\right]~, 
\end{split}\\ 
\intertext{and its trace is} 
\label{nH_tr_exp} 
  n\K^i_i &\to \frac{2d}{d-1} W  
  + \e{-2A} c +\frac12 e^i_k \partial_r e^k_i  
  - \e{-2A} e^i_k \left( \partial^k d_i + \frac{\partial_i 
  \partial^k}{\Box} c \right)~.\\ 
\intertext{The intrinsic Ricci tensor is replaced by} 
\label{R_ij_exp} 
\begin{split} 
  R_{ij} &\to \frac12 \Box e_{ij} +\frac12 e^k_l \left( \partial_i 
  \partial_k e^l_j + \partial_j \partial_k e^l_i -\partial_k 
  \partial^l e_{ij} -\partial_i \partial_j e^l_k \right) \\ 
  &\quad -\frac14 (\partial_i e^k_l)(\partial_j e^l_k)  
  +\frac12 (\partial_l e^k_i)(\partial_k e^l_j)  
  -\frac12 (\partial_l e^k_j)(\partial^l e_{ik})~, 
\end{split}\\ 
\intertext{and the Ricci scalar becomes} 
\label{R_exp} 
  R &\to - \e{-2A} \left[ e^i_j \Box e^j_i 
  + \frac34 (\partial_i e^j_k)(\partial^i e^k_j)  
  -\frac12 (\partial_i e^k_j)(\partial^j e^i_k) \right]~. 
\end{align} 
 
We finish this subsection by making two remarks on our method.  
First, the substitutions \eqref{field_subs} can 
also be interpreted as a choice of gauge, namely $X=0$.  
Second, it is straightforward to extend the present analysis to the 
case with more than one scalar. For scalars $\phi^I$ that have a 
canonical kinetic term one can define the gauge invariants $a^I$ by 
replacing in \eqref{A} $W_\phi$ by $W_I= \partial W/ \partial 
\phi^I$. In particular, inert scalars are gauge invariant up to 
$\Of$. 
 
\section{Field Equations} 
\label{field_eq} 
Let us now consider in detail the equations governing 
the dynamics of the fluctuations of an active scalar. Active scalars 
mix 
with the metric fluctuations even at the linearized level, and a lot 
of effort was devoted in previous work to resolve this mixing 
\cite{DeWolfe:2000xi, Arutyunov:2000rq, Bianchi:2000sm, Muck:2001cy, 
  Bianchi:2001de}.   
As we shall see, in our gauge invariant approach, or, 
equivalently, in the gauge $X=0$, this is much simpler. We will 
obtain 
a second order ODE for the active scalar fluctuation in a rather 
straightforward fashion.  
 
We start with the equation of motion for the 
scalar\footnote{The equation for the scalar follows from  
  Einstein's equations. However, we need not consider the 
  components $E_{ij}$, if we are not interested in the fluctuations 
  $e^i_j$.},  
\eqref{scalar_eq}, which takes the form 
\begin{multline} 
\label{eqmot_scal} 
  \left[ \partial_r^2 - 2 n^i \partial_r \partial_i +n^i n^j 
  \nabla_i \partial_j + n^2 \nabla^2 -\left( n \K^i_i +\partial_r \ln 
n 
  -n^i \partial_i \ln n \right) \partial_r \right.\\ 
  \left. - \left( \partial_r n^k -n^i \nabla_i n^k - n \nabla^k n- 
n^k 
  \partial_r \ln n + n^k n^i \partial_i \ln n - n^k n\K^i_i \right) 
  \partial_k \right] \phi - n^2 \frac{\partial V}{\partial \phi} =0~. 
\end{multline} 
We have used the expressions listed in Sec.~\ref{geom} for the 
bulk connections. 
 
After expanding \eqref{eqmot_scal} to second order and using the 
substitution rule \eqref{field_subs}, we obtain 
\begin{equation} 
\label{eq_a} 
  \left( \partial_r^2 -\frac{2d}{d-1} W \partial_r +\e{-2A} \Box 
  -V_{\phi\phi} \right) a -W_\phi \e{-2A} c -W_\phi \partial_r b 
  -2 V_\phi b = J_a~, 
\end{equation} 
where the quadratic source $J_a$ is given by  
\begin{equation} 
\label{J_a} 
\begin{split} 
  J_a &=  
  \frac12 V_{\phi\phi\phi} a^2 + V_\phi b^2 + 2 V_{\phi\phi} ab 
  - W_\phi b \partial_r b + (\partial_r a)(\partial_r b)  
  + \frac12 W_\phi e^i_j \partial_r e^j_i \\ 
  &\quad +\e{-2A} \left[ -2 b\Box a -(\partial^i b)(\partial_i a)  
  + c \partial_r a + 2\left( d^i +\frac{\partial^i c}{\Box} \right)  
  \partial_i \partial_r a  \right.\\ 
  &\quad - W_\phi \left(d^i +\frac{\partial^i c}{\Box} \right) 
\partial_i b  
  + \left( \partial_r  d^i + \partial_r \frac{\partial^i c}{\Box} 
\right) \partial_i a  
  - 2\frac{d-2}{d-1}W \left(d^i +\frac{\partial^i c}{\Box} \right) 
  \partial_i a \\ 
  &\left. \phantom{\frac12}  
  + e^i_j \left(\partial_i \partial^j a -W_\phi \partial_i d^j 
-W_\phi  
  \frac{\partial_i  \partial^j}{\Box} c\right) \right]~. 
\end{split}  
\end{equation} 
 
Next, we turn to Einstein's equations, which we need to consider 
only the 
normal and mixed components of. These are easily obtained multiplying 
\eqref{Einstein} by $N^\mu N^\nu-g^{ij}X_i^\mu X_j^\nu$ and $N^\mu 
X_i^\nu$, respectively,  
using the geometrical relations of Sec.~\ref{geom}.  
For the normal components we find  
\begin{multline} 
\label{Einst_NN} 
  (n\K^i_j)(n\K^j_i) -(n\K^i_i)(n\K^j_j) -n^2 R  
  -2n^2 g^{ij} (\partial_i \phi)(\partial_j \phi) -4 n^2 V \\    
  + 2 (\partial_r \phi)^2 -4 n^i (\partial_i \phi)(\partial_r \phi)
  + 2 n^i n^j (\partial_i \phi)(\partial_j \phi) =0~. 
\end{multline} 
Expanding to second order and replacing the fields by means of 
\eqref{field_subs} we obtain  
\begin{equation} 
\label{eq_c} 
  -4W \e{-2A} c + 4 W_\phi \partial_r a -4 V_\phi a -8 V b =J_c~, 
\end{equation} 
with the quadratic source term $J_c$ given by  
\begin{equation} 
\label{J_c} 
\begin{split} 
  J_c &= 4V b^2 +8 V_\phi ab +2 V_{\phi\phi} a^2 - 2 (\partial_r 
a)^2  
  + (\e{-2A} c)^2 + 2 \e{-2A} (\partial^i a)(\partial_i a) \\ 
  &\quad 
  + 4 W_\phi \e{-2A} \left(d^i +\frac{\partial^i c}{\Box} \right) 
  \partial_i a 
  +2 W e^i_j \partial_r e^j_i  
  -4 W \e{-2A} e^i_j \partial_i  
  \left(d^j +\frac{\partial^j c}{\Box} \right) \\ 
  &\quad 
  -\frac14 (\partial_r e^i_j)(\partial_r e^j_i)  
  +\e{-2A} \left( \partial_i d^j +\frac{\partial_i\partial^j}{\Box}
  c\right) \partial_r e^i_j \\ 
  &\quad -\e{-4A} \left[ \frac12 (\partial_i d^j)(\partial^i d_j)
  +\frac12 (\partial_i d^j)(\partial_j d^i) +2 (\partial_i d^j)  
  \frac{\partial^i\partial_j}{\Box} c +
  \left(\frac{\partial_i \partial^j}{\Box} c \right) 
  \left(\frac{\partial^i \partial_j}{\Box} c \right)  \right] \\ 
  &\quad -\e{-2A} \left[e^i_j  \Box e^j_i 
  +\frac34 (\partial_i e^j_k)(\partial^i e_j^k)  
  -\frac12 (\partial_i e^j_k)(\partial^k e^i_j) 
  \right]~. 
\end{split} 
\end{equation} 
 
Similarly, the mixed components of Einstein's equations  
are rewritten as  
\begin{multline} 
\label{Einst_Ni} 
  \partial_i (n\K^j_j) - \nabla_j (n \K^j_i) -(\partial_i \ln n) 
(n\K^j_j) 
  + (\partial_j \ln n) (n\K^j_i)  
  - 2 (\partial_i \phi) (\partial_r\phi -n^j \partial_j \phi) =0~, 
\end{multline} 
which yields the equation 
\begin{equation} 
\label{eq_bd} 
  -\frac12 \e{-2A} \Box d_i -2 W \partial_i b -2W_\phi \partial_i a = 
   J_i~. 
\end{equation} 
where $J_i$ to quadratic order is given by  
\begin{equation} 
\label{J_i} 
\begin{split} 
  J_i &= - W \partial_i b^2  
  + 2 (\partial_i a)(\partial_r a) 
  + \e{-2A} (\Pi^j_i  c) (\partial_j b)
  + \frac12 (\partial_j b)(\partial_r e^j_i) 
  - \frac12 \e{-2A} \frac{\partial_j c}{\Box} \Box e^j_i\\ 
  &\quad   
  - \frac14 \partial_i \partial_r (e^j_k e^k_j)  
  + \frac12 e^j_k \partial_r \partial_j e^k_i 
  + \frac14 (\partial_i e^j_k)(\partial_r e^k_j) 
  - \frac12 \e{-2A} e^j_k \partial_j (\partial^k d_i -\partial_i 
d^k)\\ 
  &\quad 
  - \frac12 \e{-2A} (\partial_j e^k_i) (\partial^j d_k -\partial_k 
d^j) 
  - \frac12 \e{-2A} d_j \Box e^j_i 
  - \frac12 \e{-2A} (\partial_j b)(\partial^j d_i +\partial_i d^j)~. 
\end{split} 
\end{equation}

Our strategy is to solve \eqref{eq_c} and \eqref{eq_bd} for 
$b$, $c$ and $d_i$ and substitute them into \eqref{eq_a} to obtain an 
equation for $a$, which is still coupled to $e^i_j$ through the 
source 
terms. Thus, we find 
\begin{equation} 
\label{bcd_sol}  
\begin{split} 
  b &= -\frac{W_\phi^2}{W^2} \ta -\frac1{2W} \frac{\partial_i}{\Box} 
J^i~,\\ 
  \Box d_i &= -2 \e{2A} \Pi^j_i J_j~,\\ 
  \e{-2A} c &= \frac{W_\phi^2}{W^2} \partial_r \ta   
  - \frac1{4W} J_c +\frac{V}{W^2} \frac{\partial_i}{\Box} J^i~, 
\end{split} 
\end{equation} 
where we have defined $\ta = (W/W_\phi) a$ for later convenience. 
 
Substituting \eqref{bcd_sol} into \eqref{eq_a} yields 
\begin{equation} 
\label{eq_ta} 
  \left(D^2 +\e{-2A} \Box \right) \ta = 
  J_{\ta}~, 
\end{equation} 
where we have abbreviated 
\begin{equation} 
\label{D2def} 
  D^2 = \left[ \partial_r + 2 \left(W_{\phi\phi} -\frac{W_\phi^2}{W} 
  -\frac{d}{d-1}W \right) \right] \partial_r~, 
\end{equation} 
and the source term is given by 
\begin{equation} 
\label{J_ta} 
  J_{\ta} = \frac{W}{W_\phi} J_a -\frac14 J_c -\frac12 
   \left[ \partial_r   
  + 2 \left( W_{\phi\phi} -\frac{W_\phi^2}{W} 
  -\frac{d}{d-1}W \right) \right] \frac{\partial_i}{\Box} J^i~. 
\end{equation} 
As promised, we find that the equation for $\ta$ at the linearized 
level is rather simple compared to the effort needed in previous 
work. More importantly, since it is a second order ordinary 
differential equation (after going to momentum space), it is possible 
to use the standard Green's function method for going beyond the 
linearized level.  
 
\section{Correlation Functions} 
\label{corr} 
\subsection{General Considerations} 
\label{corr_gen} 
We shall now calculate correlation functions of the dual deformed 
conformal field theory. The presentation in this and 
the next subsection will be general, if not otherwise indicated. We 
start here with some general considerations including also the use of 
the Green's function method for calculating the interaction 
terms. We use the variable $\rho=\e{-2r}$ in many formulae. In  
subsection \ref{corr_OOO} the calculation of the three-point function 
of the active scalar will be presented. Subsection \ref{corr_GPPZ} is 
dedicated entirely to the results for the GPPZ flow.  
As already mentioned, useful relations for the GPPZ flow are listed in 
appendix~\ref{GPPZ_rels}.

The bulk fluctuations $h^i_j$ and $\varphi$ 
are the duals of the boundary energy momentum tensor, $T^i_j$,  
and the scalar operator $\mathcal{O}$ of conformal dimension 
$\Delta$, 
respectively. This is made explicit through the couplings of their 
boundary values to these operators, 
\begin{equation} 
\label{op_coupling} 
  \int \rmd^d x \left( \frac12 \hh^j_i T^i_j + \hp \mathcal{O} 
\right)~, 
\end{equation} 
where $\hh^i_j$ and $\hp$ are defined as the leading coefficients in 
the asymptotic expansions  
\begin{align} 
\label{h_asympt} 
  h^i_j(x,\rho)   &= \hh^i_j(x) + \cdots + \rh^i_j(x) \rho^{d/2} 
+\cdots~,\\ 
\label{phi_asympt} 
  \varphi(x,\rho) &= \hp(x) \rho^{(d-\Delta)/2} + \cdots  
   + \rp(x) \rho^{\Delta/2} +\cdots~. 
\end{align} 
Moreover, $\rh^i_j$ and $\rp$ are the leading coefficients of the 
sub-leading 
series and are called the responses \cite{Muck:2001cy}. We have not 
written the sub-leading terms, which can also include logarithms.   
The response functions determine the \emph{exact} 
one-point functions of the corresponding dual operators.  
More precisely, in the transversal gauge, $\nu=\nu^i=0$, the exact
one-point function $\vev{\mathcal{O}}$ becomes  
\cite{Bianchi:2001kw,Martelli:2002sp} 
\begin{equation} 
\label{O_exact} 
  \vev{\mathcal{O}} = (2\Delta-d) \rp + \text{contact 
terms}~. 
\end{equation} 
where the contact terms are finite but in principle scheme dependent.
However, we can express $\vev{\mathcal{O}}$ in a manifestly gauge
invariant form after observing that $\rp$ is equal to $\ra$ plus a
(gauge dependent) function of the sources, so that \eqref{O_exact} simply 
becomes
\begin{equation} 
\label{O_exact_a} 
  \vev{\mathcal{O}} = (2\Delta-d) \ra + \text{contact 
terms}~. 
\end{equation} 
We shall not be concerned with the finite scheme dependent contact terms 
that do not affect on-shell quantities, but 
play a crucial role in the consistency of the subtraction procedure 
\cite{Bianchi:2001kw}
and contribute to certain sum rules \cite{Anselmi:2002fk}. 
 
Holographic renormalization yields the correct (anomalous) Ward
identities, which 
impose restrictions on the exact one-point function $\langle 
T^i_j\rangle$  
\cite{Bianchi:2001kw,Martelli:2002sp}, 
\begin{align} 
\label{Ward_diff} 
  \nabla_i \vev{T^i_j} + \nabla_j \hat{\phi} \vev{\mathcal{O}} 
&=0~,\\ 
\label{Ward_scale} 
  \vev{T} +(d-\Delta) \hat{\phi} \vev{\mathcal{O}} &=\mathcal{A}~. 
\end{align} 
Here, $\mathcal{A}$ denotes the conformal anomaly, and $\hat{\phi}$ 
includes also the background source, whereas $\hat{\varphi}$ only
denotes the fluctuation source. Hence, we can quite generally  
restrict the analysis of correlation functions to those  
containing $\mathcal{O}$ and the traceless transversal part of 
$T^i_j$.  
 
Eqns.\  \eqref{Ward_diff} and \eqref{Ward_scale} give rise to 
the following operator identities, which are valid in all 
(\emph{non-local}) 
correlation functions at distinct insertion points \footnote{There are
  also local terms in the correlation functions, which cannot be
  described by these operator identities. For example, from
  \eqref{Ward_scale} follows  
  \[ \vev{T(z)\mathcal{O}(x) \mathcal{O}(y)} =  
     \beta \vev{\mathcal{O}(z) \mathcal{O}(x) \mathcal{O}(y)} 
     +\left[ \delta(z-x) +\delta(z-y) \right]  
       \vev{\mathcal{O}(x) \mathcal{O}(y)}  
     +\frac{\partial}{\partial z^i} \delta(z-x)  
      \frac{\partial}{\partial z^i} \delta(z-y)~. \] 
  The last term stems from the anomaly.} 
\begin{equation} 
\label{op_id} 
  \partial_i T^i_j =0~,\qquad  T = \beta \mathcal{O}~. 
\end{equation}  
For the GPPZ flow, where $d=4$, $\Delta=3$, and
$\hat{\bar{\phi}}=\sqrt{3}$, we find $\beta=-\sqrt{3}$. 
These operator identities can also be understood from our gauge  
invariant approach. Substituting the decomposition \eqref{hsplit} 
into 
\eqref{op_coupling}, we see that $\epsilon^j$ is the source of 
$\partial_i T^i_j$. However, $\epsilon^j$ appears in the gauge 
invariant fields only with an $r$-derivative, so that its source, 
which appears in the constant leading term, is irrelevant for the 
bulk 
dynamics. Consequently, the dual operator vanishes. 
Similarly, up to the numerical constant $2(d-1)$, $\hh$ is the source of 
$T$, and from the definition of the invariant $a$ we see 
that the corresponding source is  
\begin{equation}  
  \ha = \hp -\frac{(\Delta-d)}{2(d-1)}\hat{\bar{\phi}} \hh~, 
\end{equation} 
from which the second identity in \eqref{op_id} follows immediately.

Let us now consider in more detail the exact holographic one-point 
function 
$\vev{\mathcal{O}}$ in the presence of sources. As we are not
interested in finite scheme dependent contact terms, it suffices to
calculate the 
response $\ra$. To this end we observe that $a$ satisfies an equation
of the form [cf.\ \eqref{eq_ta}]  
\begin{equation} 
\label{eq_a_form} 
  \left( \tilde{\nabla}^2 - M^2 \right) a = \frac{W_\phi}{W} 
J_{\ta}~, 
\end{equation} 
where $M^2$ is an effective mass term, and $\tilde{\nabla}$ now 
denotes 
the background covariant derivative. Thus, after defining a 
covariant Green's function by  
\begin{equation} 
\label{Green_def} 
  \left( \tilde{\nabla}^2 - M^2 \right) G(z,z') = 
  \frac{\delta(z-z')}{\sqrt{\tg(z)}}~, 
\end{equation} 
the general solution for $a$ has the form  
\begin{equation} 
\label{a_sol_form} 
  a(z) = \int \rmd^d y\, K(z,y) \ha(y) + \int \rmd^{d+1} z'\, 
  \sqrt{\tg(z')} G(z,z') \frac{W_\phi(z')}{W(z')} J_{\ta}(z')~. 
\end{equation} 
Here, $z$ is a short notation for the variables $(\rho,x)$, and $x$ 
and 
$y$ are boundary coordinates. Notice 
that the bulk integral is cut off at $\rho'=\varepsilon$, and 
also that $\rho\ge \varepsilon$, because of the regularization 
procedure.  
Moreover, $K(z,y)$ denotes the bulk-to-boundary propagator.  
 
We are interested in the near-boundary behaviour of $a$, 
and it is very helpful that in asymptotically AdS spaces the Green's 
function asymptotically behaves as \cite{Muck:1999kk} 
\begin{equation} 
\label{Green_asympt} 
  G(z,z') \approx - \frac{\rho^{\Delta/2}}{2\Delta-d} K(x,z') 
+\cdots~. 
\end{equation}  
Setting also $\rho=\epsilon$, this yields  
\begin{equation} 
\label{a_asympt} 
  a(\varepsilon,x) \approx \int \rmd^d y\, K(\varepsilon,x;y) \ha(y)  
  - \frac{\varepsilon^{\Delta/2}}{2\Delta-d} \int\limits_\varepsilon 
\frac{\rmd \rho'}{2\rho'}  
  \int \rmd^d y\, \e{dA(\rho')}   
  K(x;\rho',y) \frac{W_\phi(\rho')}{W(\rho')} J_{\ta}(\rho',y)~. 
\end{equation} 
It is more useful to consider this expression in momentum space, 
where 
we can use momentum conservation in the propagators, 
$K(\rho,p;q)=K_p(\rho) \delta(p+q)$. We also introduce the 
bulk-to-boundary propagator for the field $\ta$, defined by $\tK 
=(W/W_\phi) K$. Thus, \eqref{a_asympt} becomes 
\begin{equation} 
\label{a_asympt2} 
  a(\varepsilon,p) \approx \frac{W_\phi(\varepsilon)}{W(\varepsilon)} 
\tK_p(\varepsilon) \ha(p)  
  - \frac{\varepsilon^{\Delta/2}}{2\Delta-d} \int\limits_\varepsilon 
\frac{\rmd \rho}{2\rho} \e{dA(\rho)}   
  \tK_p(\rho) \left[\frac{W_\phi(\rho)}{W(\rho)}\right]^2  
  J_{\ta}(\rho,p)~. 
\end{equation} 
 
The two-point function $\vev{\mathcal{O}\mathcal{O}}$ can be read off 
easily from the asymptotic behaviour of the bulk-to-boundary 
propagator, $\tK_p$.  
Were the integral in \eqref{a_asympt2} finite for 
$\varepsilon\to 0$ (removal of the cut-off), it would 
directly represent the contribution to the response $\ra$ stemming 
from the 
interactions, \ie also the terms needed for the three-point 
functions.  
However, one should in general expect the integral to be 
divergent. This does not cause problems, because the divergences can  
be understood and even predicted by considering the counter terms  
in holographic renormalization and are thus easily removed. 
(Remember that holographic renormalization ensures that 
the exact one-point function is finite.) The counter terms are
obtained quite easily using the Hamilton-Jacobi approach to
holographic normalization \cite{Martelli:2002sp}.  
In the GPPZ flow, such terms do not appear  
with the active scalar, because the generic $\phi^4$ logarithmic 
counter term is absent. In contrast, we expect logarithmic 
divergences from 
the $(e^i_j)^2$ terms in $J_{\ta}$, since there is a $\phi^2 R$ 
logarithmic counter term, and also in the mixed  
active-inert scalar sector because there are 
$\phi^2 \sigma^2$ logarithmic counter terms \cite{Bianchi:2003bd}.

\subsection{The Holographic Three-Point Function $\vev{\mathcal{O}\mathcal{O}
\mathcal{O}}$} \label{corr_OOO} 

In this subsection, we give a general, but formal expression for the 
three-point function $\vev{\mathcal{O}\mathcal{O}\mathcal{O}}$, and 
we 
demonstrate that it is Bose symmetric. This is a useful check, 
because 
we do not differentiate three times with respect to the source  
a Lagrangian for the field $\ta$ alone. 
 
The relevant interaction terms are obtained by inserting the linear 
solutions for $b$ and $c$ from \eqref{bcd_sol} into \eqref{J_ta} 
dropping $d_i$ and $e^i_j$. Thus, we find  
\begin{align} 
\notag  J_{\ta} &= \frac14 \left( \frac{W_\phi}{W} \right)^4  
  \left[ 2 \frac{\partial^i\partial_j}{\Box} 
  \left( \ta' \Pi^j_i \ta' \right) - (\Pi^j_i \ta') 
  \frac{\partial^i\partial_j}{\Box}  \ta' \right]  
  + 2 \left( \frac{W_\phi}{W} \right)^2  
  \frac{\partial^i \ta'}{\Box} \partial_i \ta'\\ 
\notag  &\quad +\frac12 \left[\partial_r \left( \frac{W_\phi}{W}
  \right)^4 \right]  
  \frac{\partial^i\partial_j}{\Box} \left( \ta \Pi^j_i \ta' 
  \right)  
  + \left[\partial_r \left( \frac{W_\phi}{W} \right)^2 \right]  
  \left[ \frac{\partial^i \ta'}{\Box}  \partial_i  \ta 
  - \frac{\partial^i}{\Box} \left( \ta' \partial_i \ta  
  \right) - \ta'  \ta \right] \\ 
\notag  &\quad -\frac12 \left( \frac{W_\phi}{W} \right)^4 \e{-2A} 
  \frac{\partial^i\partial_j}{\Box} \left(\ta \Box \Pi^j_i \ta 
  \right)  
  +  \frac12 \left[D^2 \left( \frac{W_{\phi\phi}}{W} - 
  \frac{W_\phi^2}{W^2} \right)\right] \ta^2\\
\label{J_ta2} 
  &\quad + \left( \frac{W_\phi}{W} \right)^2 \e{-2A} \left[ 2\ta 
\Box \ta +  
  \frac{\partial^i}{\Box} \left[ (\partial_i \ta)(\Box \ta) \right] 
  -\frac12 (\partial^i \ta)(\partial_i \ta) \right]~,
\end{align} 
where $D^2$ is the second order differential operator defined in
\eqref{D2def}, and we have abbreviated $\ta'=\partial_r \ta$.

The three-point function $\vev{\mathcal{O}\mathcal{O}\mathcal{O}}$ is
formally given by the integral in \eqref{a_asympt2}, where we should
substitute the first order solutions of $\ta$ into $J_{\ta}$. However,
in this form it is not evident that the final expression will be Bose
symmetric. In fact, we expect the integrand to be Bose symmetric up to
total derivative terms, which then must vanish in the $\varepsilon\to
0$ limit. In order to facilitate the integrations by parts, it is
helpful to perform first a field redefinition that removes from  $J_{\ta}$ 
the terms of the form $(\ta')^2$ in the 
first line of \eqref{J_ta2},
\cite{Lee:1998bx}. Hence, we perform the replacement 
\begin{equation} 
\label{a_redef} 
 \ta \to \ta +\frac18 \left( \frac{W_\phi}{W} \right)^4  
  \left[ 2 \frac{\partial^i\partial_j}{\Box} \left( \ta 
  \Pi^j_i \ta \right) - (\Pi^j_i \ta) 
  \frac{\partial^i\partial_j}{\Box} \ta \right]  
  + \left( \frac{W_\phi}{W} \right)^2  \frac{\partial^i \ta}{\Box} 
  \partial_i \ta~. 
\end{equation} 
After this field redefinition the source in \eqref{eq_ta} becomes 
\begin{align} 
\notag 
  J_{\ta} &= \frac12 \left[\partial_r \left( \frac{W_\phi}{W}
  \right)^4 \right]  
  \left[ \frac{\partial^i\partial_j}{\Box} \left( \ta' 
         \frac{\partial^j\partial_i}{\Box} \ta \right) 
     -   \left(\frac{\partial^i\partial_j}{\Box} \ta'\right) 
         \left(\frac{\partial^j\partial_i}{\Box} \ta \right) 
\right]\\ 
\notag &\quad 
  - \left[ \partial_r \left( \frac{W_\phi}{W} \right)^2\right]  
  \left[ \frac{\partial^i \ta'}{\Box} \partial_i \ta 
  +2 \frac{\partial^i \ta}{\Box} \partial_i \ta'  
  + \frac{\partial^i}{\Box} \left( \ta' \partial_i \ta \right) 
  + \ta \ta' \right] \\ 
\notag &\quad  
  - \frac12 \left( \frac{W_\phi}{W} \right)^4 \e{-2A} 
  \left[ \ta \Box \ta - \frac{\partial^i\partial_j}{\Box} \left( \ta 
  \partial_i \partial^j \ta \right) 
  +\frac12 (\partial_i \ta)(\partial^i \ta) \right.\\ 
\notag &\quad \left.  
  - \frac{\partial^i\partial_j}{\Box} \left( (\partial_k \ta) 
  \frac{\partial_i \partial^j \partial^k}{\Box} \ta \right) 
  + \frac12 \left(\frac{\partial^i \partial^j \partial^k}{\Box} 
\ta\right)  
            \left(\frac{\partial_i \partial_j \partial_k}{\Box}
  \ta\right) \right] \\
\notag &\quad 
  + \left( \frac{W_\phi}{W} \right)^2 \e{-2A} 
  \left\{ 2\ta \Box \ta + \frac{\partial^i}{\Box} \left[ (\partial_i 
\ta) 
  (\Box \ta) \right] -\frac12 (\partial^i\ta)(\partial_i \ta) 
  - 2 \left(\frac{\partial^i\partial_j}{\Box} \ta\right)(\partial_i 
\partial^j \ta) 
  \right\} \\ 
\notag &\quad 
  +\frac12 \left[D^2 \left( \frac{W_{\phi\phi}}{W} - 
    \frac{W_\phi^2}{W^2} \right) \right] \ta^2  
  - \left[D^2 \left( \frac{W_\phi}{W} \right)^2\right]  
  \frac{\partial^i\ta}{\Box} \partial_i \ta \\ 
\label{J_ta_new} 
  &\quad -\frac18 \left[D^2 \left( \frac{W_\phi}{W} \right)^4\right]  
   \left[ \ta^2 -2 \frac{\partial^i\partial_j}{\Box} \left( \ta 
   \frac{\partial^j\partial_i}{\Box} \ta \right) + 
   \left(\frac{\partial^i\partial_j}{\Box} \ta \right)
   \left(\frac{\partial^j\partial_i}{\Box} \ta \right)\right]~. 
\end{align} 
 
The three-point function of the active scalar 
is given by the integral in \eqref{a_asympt2},\footnote{We 
  might need to subtract divergences, if holographic renormalization 
  predicts them by the presence of the appropriate counter terms.} 
\begin{equation} 
\label{ra_2} 
  (2\Delta-d) \ra^{(2)} (p) = - \int\limits_0^\infty \rmd r\,\e{dA} 
  \left(\frac{W_\phi}{W}\right)^2  \tK_p(r) J_{\ta}(r,p)~. 
\end{equation} 
 
Differentiating $\ra^{(2)}$ twice with respect to $\ha(-p)$ 
yields the three-point function  
$\vev{\mathcal{O} \mathcal{O} \mathcal{O}}$. 
In order to do this, note that $J_{\ta}(r,p)$ is of 
the form  
\begin{equation} 
\label{Jform} 
  J_{\ta}(r,p) = \int \rmd p_2\, \rmd p_3\, \delta(p+p_2+p_3) 
  \X(p,-p_2,-p_3) \tK_2 \tK_3 \ha(-p_2) \ha(-p_3)~, 
\end{equation} 
where $\tK_2$ and $\tK_3$ stand for the bulk-to-boundary propagators 
$\tK_{p_2}(r)$ and $\tK_{p_3}(r)$, respectively, and the operator 
$\X$, 
which includes derivatives with respect to $r$ acting on $\tK_2$ 
and $\tK_3$, can be read off from \eqref{J_ta_new}. It is important 
to 
notice the minus signs of the momenta $p_2$ and $p_3$. Thus, the
three-point function we are seeking is
\begin{equation} 
\label{OOO} 
  \vev{\mathcal{O}_1 \mathcal{O}_2 \mathcal{O}_3} = - 
  \delta(p_1+p_2+p_3) \int\limits_0^\infty \rmd r\,  
  \e{dA} \left(\frac{W_\phi}{W}\right)^2 \X_{123} \tK_1 \tK_2 \tK_3~, 
\end{equation} 
where  
\begin{equation} 
\label{Xdef} 
  \X_{123} = \X(p_1,-p_2,-p_3) + \X(p_1,-p_3,-p_2)~. 
\end{equation} 
 
After reading off $\X$ from \eqref{J_ta_new} and using momentum 
conservation, 
$p_1+p_2+p_3=0$, we find 
\begin{align} 
\notag 
  \X_{123} &= \frac12 \left[\partial_r \left(
  \frac{W_\phi}{W}\right)^4 \right]  
  \left[ \frac{(p_1\cdot p_2)^2}{p_1^2p_2^2} \partial_3 +  
         \frac{(p_1\cdot p_3)^2}{p_1^2p_3^2} \partial_2 +   
         \frac{(p_2\cdot p_3)^2}{p_2^2p_3^2} \partial_1  \right.\\ 
\notag &\quad \left. 
       - \frac{(p_2\cdot p_3)^2}{p_2^2p_3^2}  
         (\partial_1 +\partial_2 +\partial_3) \right] \\ 
\notag &\quad 
  + \left[\partial_r \left( \frac{W_\phi}{W}\right)^2 \right]  
  \left[  
  p_2\cdot p_3 \left(\frac1{p_2^2}+\frac1{p_3^2}\right) \partial_1 +  
  p_1\cdot p_3 \left(\frac1{p_1^2}+\frac1{p_3^2}\right) \partial_2
  \right.\\
\notag &\quad \left. 
  + p_1\cdot p_2 \left(\frac1{p_1^2}+\frac1{p_2^2}\right) \partial_3  
  - p_2\cdot p_3 \left(\frac1{p_2^2}+\frac1{p_3^2}\right)  
  (\partial_1 +\partial_2 +\partial_3) \right] \\ 
\notag &\quad 
  +\frac12 \left( \frac{W_\phi}{W}\right)^4 \e{-2A}  
  \left[ \frac12 (p_1^2+p_2^2+p_3^2) +  
   \frac{(p_1\cdot p_2)^3}{p_1^2p_2^2} +  
   \frac{(p_1\cdot p_3)^3}{p_1^2p_3^2} +   
   \frac{(p_2\cdot p_3)^3}{p_2^2p_3^2} \right] \\ 
\notag &\quad 
  + \left( \frac{W_\phi}{W}\right)^2 \e{-2A} 
  \left[ (p_2\cdot p_3)^2 \left(\frac1{p_2^2}+\frac1{p_3^2}\right) +  
         (p_1\cdot p_3)^2 \left(\frac1{p_1^2}+\frac1{p_3^2}\right)
  \right. \\
\notag & \quad \left. 
         + (p_1\cdot p_2)^2 \left(\frac1{p_1^2}+\frac1{p_2^2}\right) 
         - 2(p_1^2+p_2^2+p_3^2) \right] \\
\notag &\quad
  +\frac14 \left[D^2 \left( \frac{W_\phi}{W}\right)^4\right]  
  \left[ \frac{(p_1\cdot p_2)^2}{p_1^2p_2^2} +  
         \frac{(p_1\cdot p_3)^2}{p_1^2p_3^2} -   
         \frac{(p_2\cdot p_3)^2}{p_2^2p_3^2} \right]\\ 
\label{X123} 
  &\quad + \left\{ D^2 \left[ \frac{W_{\phi\phi}}{W} 
  -\left(\frac{W_\phi}{W}\right)^2  
  -\frac14 \left(\frac{W_\phi}{W}\right)^4 \right] \right\} 
  -\left[ D^2  \left(\frac{W_\phi}{W}\right)^2 \right]  
   p_2\cdot p_3 \left(\frac1{p_2^2}+\frac1{p_3^2}\right)~. 
\end{align} 
The symbols $\partial_n$, $n=1,2,3$, denote the derivative with 
respect 
to $r$ acting on the bulk-to-boundary propagator $\tK_n$. As 
\begin{equation}  
  (\partial_1 +\partial_2 +\partial_3) (\tK_1 \tK_2 \tK_3) = 
\partial_r (\tK_1 
  \tK_2 \tK_3)~, 
\end{equation} 
we can integrate these two terms of \eqref{X123} by parts in the 
integral in \eqref{OOO} exploiting also the identity 
\begin{equation} 
\label{D2ident} 
  \partial_r \left[\e{dA} \left(\frac{W_\phi}{W}\right)^2 \partial_r 
  \right] = \e{dA} \left(\frac{W_\phi}{W}\right)^2 D^2~. 
\end{equation} 
Hence, the last term in \eqref{X123} is cancelled, and the minus sign 
of the last term on the penultimate line is reversed, rendering the 
final result totally symmetric in the indices 1, 2 and 3. This is the 
proof of Bose symmetry we wanted.  
Thus, the final result is  
\begin{align} 
\notag 
  \X_{123} &= \frac12 \left[\partial_r \left(
  \frac{W_\phi}{W}\right)^4 \right] 
  \frac{(p_1\cdot p_2)^2}{p_1^2p_2^2} \partial_3   
  + \left[\partial_r \left( \frac{W_\phi}{W}\right)^2 \right] 
  p_1\cdot p_2 \left(\frac1{p_1^2}+\frac1{p_2^2}\right) \partial_3 \\ 
\notag &\quad  
  +\frac12 \left( \frac{W_\phi}{W}\right)^4 \e{-2A}  
  \left[ \frac12 p_1^2 + \frac{(p_1\cdot p_2)^3}{p_1^2p_2^2} \right]  
  + \left( \frac{W_\phi}{W}\right)^2 \e{-2A} 
  \left[ (p_1\cdot p_2)^2 \left(\frac1{p_1^2}+\frac1{p_2^2}\right) 
  -2p_1^2 \right] \\ 
\label{X123final} &\quad  
  +\frac14 \left[D^2 \left( \frac{W_\phi}{W}\right)^4 \right]  
   \frac{(p_1\cdot p_2)^2}{p_1^2p_2^2}  
  +\frac13 \left\{D^2 \left[ \frac{W_{\phi\phi}}{W} 
  -\left(\frac{W_\phi}{W}\right)^2  
  -\frac14 \left(\frac{W_\phi}{W}\right)^4 \right] \right\}
  +\text{cyclic}~.  
\end{align}

In the above analysis we have not really been careful dropping the 
boundary terms in the integration by parts  
in the step from \eqref{X123} to \eqref{X123final} and 
using the field redefinition \eqref{a_redef}, both of which might 
contribute terms to the three-point function that are not Bose 
symmetric. Thus, it remains to check that these contributions either
cancel or vanish in the limit $\varepsilon\to0$. 
We shall show this explicitly for the GPPZ 
flow in the next subsection, but the argument extends easily to the 
general case.

\subsection{Correlation Functions in the GPPZ Flow} 
\label{corr_GPPZ} 
We shall now apply our results to the GPPZ flow, where it is
conventional to use the variable $u=1-\rho=1-\e{-2r}$. 
Useful identities are listed in appendix~\ref{GPPZ_rels}. For
completeness, we also calculate the two-point function using our
simpler linear equation, although the result has been known for some
time. We would like to mention that we have performed the calculations
for dimensionless $p$. In order to restore the proper dimensions, one
must replace $p$ by $pL$ everywhere, where $L$ is the radius of 
curvature 
of the asymptotic AdS region ($L^4=4\pi g_s N {\alpha'}^2$), so that
$\mathcal{O}(p)\to \mathcal{O}(pL) =\mathcal{O}(p)/L$. 
($\mathcal{O}(p)$ has dimension $-1$, corresponding to
$\mathcal{O}(x)$ of dimension $3$.) Furthermore, the results should be
multiplied by the numerical factor $[N^2/(2\pi^2)]\times (2\pi)^4$,
where the $(2\pi)^4$ stems from our convention for the
$\delta$-function in momentum space.

The equation of motion \eqref{eq_ta} for $\ta$ becomes (in momentum 
space)  
\begin{equation} 
\label{GPPZ_eq_ta} 
  \left[ u(1-u) \partial_u^2 +(2-2u) \partial_u -\frac{p^2}{4} 
\right] 
  \ta = \frac{u}{4(1-u)} J_{\ta}~. 
\end{equation} 
The associated homogeneous equation is a hypergeometric equation, whose solution,  
which 
is regular for $u=0$, is readily found. We have to be somewhat 
careful 
with the normalization, because the expansion \eqref{phi_asympt} 
should hold also for $K_p=(W_\phi/W)\tK_p$  
(in particular with a factor $1$ in the 
leading term). Hence, we find the bulk-to-boundary propagator  
\begin{equation} 
\label{GPPZ_a1_sol} 
  \tK_p (u) = \frac{\sqrt{3}}{2}  
    \Gamma \left(\frac{3+\alpha}2 \right)  
    \Gamma \left(\frac{3-\alpha}2 \right)  
  \F\left( \frac{1+\alpha}2, \frac{1-\alpha}2; 2; u \right)~, 
\end{equation} 
with $\alpha= \sqrt{1-p^2}$. Its asymptotic behaviour is given by 
\cite{Abramowitz} 
\begin{equation} 
\label{GPPZ_a1_asym} 
  \tK_p (u) \approx  \frac{\sqrt{3}}{2}  
  \left[ 1 + \frac{p^2}4 (1-u) \ln(1-u)  
  + \frac12 H(p) (1-u) +\cdots \right]~. 
\end{equation} 
The function $H(p)$, which is related to the two-point function by 
\begin{equation} 
\label{2pt} 
  \vev{\mathcal{O}(p)\mathcal{O}(q)}= \delta(p+q) H(p)~, 
\end{equation}  
is given by 
\begin{equation} 
\label{GPPZ_2pt} 
  H(p) = \frac{p^2}2 \left[  
     \psi \left( \frac{3+\alpha}2 \right)  
   + \psi \left( \frac{3-\alpha}2 \right) -\psi(2) -\psi(1) \right]~. 
\end{equation} 
where $\psi(z) = \Gamma'(z) / \Gamma(z)$. 
The spectrum of poles becomes clear after 
rewriting $H$ in a series representation using the formula 
\cite{Gradshteyn} 
\begin{equation} 
\label{psi_series} 
  \psi(x) -\psi(y) = \sum\limits_{k=0}^{\infty}  
  \left( \frac1{y+k} -\frac1{x+k} \right)~. 
\end{equation} 
This yields 
\begin{equation} 
\label{GPPZ_2pt_series} 
  H(p) = \frac{p^4}2 \sum\limits_{k=1}^\infty  
  \frac{2k+1}{k(k+1)[4k(k+1)+p^2]}~. 
\end{equation} 
Thus, we find particles with the masses $m^2= 4k(k+1)$, 
$k=1,2,3,\ldots$. The residues at the poles, which represent the decay
constants \cite{Peskin}, are   
\begin{equation} 
\label{twopt_res} 
  |f_k|^2 = 8k(k+1)(2k+1)~. 
\end{equation}

Let us now consider the three-point function.  
In the case of the GPPZ flow, we obtain from \eqref{OOO} and
\eqref{X123final} 
\begin{equation} 
\label{OOO_GPPZ} 
  \vev{\mathcal{O}_1 \mathcal{O}_2 \mathcal{O}_3} =  
  - \frac2{27} \delta(p_1+p_2+p_3) \int\limits_0^1 \rmd u\,  
  \Y_{123} \tK_1 \tK_2 \tK_3~,
\end{equation} 
where
\begin{align} 
\notag 
  \Y_{123} &= \frac{9 u^2}{(1-u)^2} \X_{123} \\
\notag 
  &=-64 u^2 (1-u)  
  \frac{(p_1\cdot p_2)^2}{p_1^2p_2^2} \partial_3   
  - 48 u^2  
  p_1\cdot p_2 \left(\frac1{p_1^2}+\frac1{p_2^2}\right) \partial_3 \\ 
\notag &\quad  
  +8 u(1-u)   
  \left[ \frac14 (p_1^2+p_2^2) 
  + \frac{(p_1\cdot p_2)^3}{p_1^2p_2^2} \right]  
  + 12 u  
  \left[ (p_1\cdot p_2)^2 \left(\frac1{p_1^2}+\frac1{p_2^2}\right) 
  -(p_1^2+p_2^2) \right] \\ 
\label{X123_GPPZ} &\quad  
  + 32 u(3u-2)  
   \frac{(p_1\cdot p_2)^2}{p_1^2p_2^2}  
  + 16 u \left[2(1-u) +\frac13 \right] 
  +\text{cyclic}~. 
\end{align} 
Here, $\partial_n$ denotes the derivative with respect to $u$ acting 
on 
$\tK_n$. It is reassuring that $\Y_{123}$ is finite for $u\to1$. 
 
As anticipated in subsection \ref{corr_OOO}, we need to check that 
the 
field redefinition and the integration by parts used to obtain the 
final result do not spoil it.  
On the one hand, the total derivative terms discarded in the step 
from 
\eqref{X123} to \eqref{X123final} are  
\begin{equation} 
\label{totder_terms} 
  \e{4A} \left(\frac{W_\phi}{W}\right)^2 \tK_1 \tK_2 \tK_3 
  \delta(p_1+p_2+p_3) \partial_r \left[ \frac12 
  \left(\frac{W_\phi}{W}\right)^4   
  \frac{(p_2\cdot p_3)^2}{p_2^2p_3^2}  
  + \left(\frac{W_\phi}{W}\right)^2 p_2 \cdot p_3 \left( 
  \frac1{p_2^2}+ \frac1{p_3^2} \right) \right]~, 
\end{equation} 
where we must take the limit $r\to\infty$. After going to the $u$ 
variable using the relations of appendix~\ref{GPPZ_rels}, as well as 
$\tK_n = \sqrt{3}/2+\cdots$ from  \eqref{GPPZ_a1_asym}, we obtain   
\begin{equation} 
\label{totder_terms2} 
  -\frac4{\sqrt{3}} p_2 \cdot p_3  
  \left( \frac1{p_2^2}+ \frac1{p_3^2} \right) \delta(p_1+p_2+p_3)~. 
\end{equation} 
On the other hand, the contribution from the field redefinition is 
found by differentiating the quadratic terms on the right 
hand side of \eqref{a_redef} twice with respect to the source 
$\ha(-p)$. This yields 
\begin{multline} 
\label{redef_terms} 
  2 \left[ \frac14 \left(\frac{W_\phi}{W}\right)^4   
  \left( 1- \frac{(p_1\cdot p_2)^2}{p_1^2p_2^2} -  
            \frac{(p_1\cdot p_3)^2}{p_1^2p_3^2} + 
            \frac{(p_2\cdot p_3)^2}{p_2^2p_3^2} \right) \right.\\ 
  \left. +  \left(\frac{W_\phi}{W}\right)^2  p_2 \cdot p_3  
  \left( \frac1{p_2^2}+ \frac1{p_3^2} \right) \right] \tK_2 \tK_3 
  \delta(p_1+p_2+p_3)  \\  
  = \frac{W}{W_\phi} \left[  
  (1-u)^{3/2}\frac{4}{\sqrt{3}} p_2 \cdot p_3  
  \left( \frac1{p_2^2}+ \frac1{p_3^2} \right) 
  +\mathcal{O}((1-u)^{5/2}) \right]\delta(p_1+p_2+p_3)~. 
\end{multline} 
We have factored out $W/W_\phi$ in order to put in
evidence the contribution to the response $\ra$. Obviously, it 
cancels 
the total derivative term \eqref{totder_terms2}.

In order to find the irreducible trilinear couplings of on-shell 
states, we need 
to amputate the legs of the three-point 
function \cite{Peskin}. This is done best by rewriting the 
bulk-to-boundary 
propagator \eqref{GPPZ_a1_sol} around an on-shell pole as  
\begin{align} 
\notag  
  \tK_p(u) &= \frac{\sqrt{3}}2  
           \Gamma \left(\frac{1+\alpha}2 \right)   
           \Gamma \left(\frac{1-\alpha}2 \right) (1-u) 
     \frac{\rmd}{\rmd u} 
  \F\left(\frac{1+\alpha}2,\frac{1-\alpha}2;1;u \right)\\ 
\label{bbprop_pole} 
  &= \frac{|f_k|}{p^2+4k(k+1)} 
  \left[\sqrt{\frac{3(2k+1)}{2k(k+1)}} 
  (1-u)  \frac{\rmd}{\rmd u} \LeP_k(2u-1)\right]  +\text{regular}~, 
\end{align} 
where $f_k$ is the decay constant defined in \eqref{twopt_res}, and
$\LeP_k$ denotes the Legendre polynomial of degree $k$, which
satisfies the useful formula 
\begin{equation} 
\label{LegPol} 
\LeP_k(2u-1) = \sum_{n=0}^{k}\binom{k}{n}^2 u^{n} (u-1)^{k-n}~.
\end{equation} 
Thus, we can write the three-point function \eqref{OOO_GPPZ} close to 
the on-shell poles as 
\begin{equation} 
\label{OOO_poles} 
  \vev{\mathcal{O}_1 \mathcal{O}_2 \mathcal{O}_3} = 
  \delta(p_1+p_2+p_3) \prod\limits_{i=1}^3 \left[ 
    \frac{|f_{k_i}|}{p_i^2 +4k_i(k_i+1)} \right] 
  \mathcal{M}+\text{regular}~,\ 
\end{equation} 
where $\mathcal{M}$ denotes the amplitude 
\begin{equation} 
\label{Mdef} 
  \mathcal{M} =  -\frac19 \sqrt{\frac32}  
  \prod\limits_{i=1}^3 \sqrt{\frac{2k_i+1}{k_i(k_i+1)}}  
  \int\limits_0^1 \rmd u\, \Y_{123} F_1 F_2 F_3~, 
\end{equation} 
with 
\begin{equation} 
  F_i(u) = (1-u) \frac{\rmd}{\rmd u} \LeP_{k_i} (2u-1)~. 
\end{equation} 
Moreover, one should substitute $p_1\cdot p_2 = (p_3^2 
-p_1^2-p_2^2)/2$ and cyclic, with $p_i^2 =-4k_i(k_i+1)$, into
\eqref{X123_GPPZ}. 

The integral in \eqref{Mdef} is elementary for all $k_i$, but, 
contrary to the simpler case of the trilinear couplings of two inert 
scalars to the active scalar \cite{Bianchi:2003bd}, the final result is not 
particularly illuminating, and we do not display it here. We would only 
like to mention that it satisfies a sort of `triangular inequality' as a 
consequence of the orthogonality of the Legendre polynomials 
$\LeP_{k_i} (2u-1)$ in the interval $u\in[0,1]$. The $1/p^2$ terms in
the operator $\Y_{123}$ defined in \eqref{X123_GPPZ} are not
dangerous, because the spectrum contains no massless states. 
Finally, \eqref{OOO_poles} has the correct (mass) dimension and large $N$ 
suppression.

\section{Conclusions and Outlook} 
\label{conc} 
In this paper, we have developed and used a gauge invariant approach 
for the analysis of the equations governing the dynamics of active 
scalar fluctuations in RG flows. This approach has enabled us 
to arrive at a second order ODE for the active scalar in a rather 
simple fashion, which in turn made it possible to use the Green's 
function method to deal with the quadratic interaction terms. 
Thus, we have taken another step beyond the tradition 
and filled a gap for active scalar operators, whose 
three-point function has not been analyzed in \cite{Bianchi:2003bd}. 
  
As an application, we derived an explicitly Bose symmetric formula 
for the 
three-point function $\vev{\mathcal{O}\mathcal{O}\mathcal{O}}$. To 
arrive at the Bose symmetric expression a field redefinition removing 
the interaction terms with two $r$-derivatives was instrumental. 
It was then necessary to integrate by parts the integral for 
the three-point function, with the boundary terms cancelling the 
contributions from the field redefinition. Other three-point 
functions 
can be calculated in the same 
fashion, which we leave as a future project. We anticipate that the 
use of the gauge invariant approach simplifies matters 
significantly.  
It would also be interesting to cross-check three-point functions 
like 
$\langle T^i_j\mathcal{O}\mathcal{O}\rangle$ calculating them in two 
different 
ways (in this case, using the Green's function for $a$ and $e^i_j$).  

As mentioned after \eqref{O_exact_a}, we have not been concerned about
the scheme dependent local terms. They are interesting, because
some central charges and sum rules may depend on them
\cite{Erdmenger:2001ja, Anselmi:2002fk, Cappelli:2000dv,
  Cappelli:2001pz}. Moreover,
in the supersymmetric renormalization
scheme the finite counter terms without derivatives are uniquely
determined. It is not clear to us whether it also determines the
other finite counter terms. Hence, a deeper investigation of this
matter is advisable. 
  
We have applied our final result to the GPPZ flow, which is 
particularly simple. As the bulk-to-boundary propagator of on-shell
states is a polynomial in this case,  
the final integral can be carried out explicitly, although it is very 
difficult to give a general formula for arbitrary masses. This can 
be used to extract, \eg the decay rates for the various decay 
channels. 
A thorough analysis involving also the states arising from the inert 
scalars and the energy-momentum tensor looks feasible.   
Moreover, we leave it to the future to apply the result to the Coulomb 
branch flow.  
 
\acknowledgments
The work of M.~B.\ and M.~P.\ was supported in part by INFN, by the EEC
contracts HPRN-CT-2000-00122 and HPRN-CT-2000-00148, and in part by the
INTAS contract 99-1-590.
W.~M.\ has been financially supported by INFN and the European 
Commission, project HPRN-CT-2000-00131.

\begin{appendix} 

\section{Useful Relations for the GPPZ Flow}
\label{GPPZ_rels}
We summarize here a number of relations for the GPPZ background. For
simplicity, we set the asymptotically AdS length scale to unity, \ie
$L=1$.

The superpotential for the active scalar is 
\begin{equation}
\label{GPPZ_W}
  W = - \frac34 \left( \cosh \frac{2\phi}{\sqrt{3}} +1 \right)~.
\end{equation}
This potential leads to a significant simplification of
its various derivatives, because of the identities
\begin{equation}
\label{GPPZ_Wphiphi}
  W_{\phi\phi} = \frac43 W + 1~,\qquad 
  \frac{W_\phi^2}{W} = \frac43 W +2~.
\end{equation}
Integrating the background equations \eqref{background} yields
\begin{equation}
\label{GPPZ_background}
  \e{2\bp/\sqrt{3}} = \frac{1+\e{-r}}{1-\e{-r}}~,\qquad
  \e{2A} = \e{2r} - 1~.
\end{equation}
From \eqref{GPPZ_background} we easily find the background source,
\begin{equation}
\label{GPPZ_b_source}
  \hat{\bar{\phi}} = \sqrt{3}~.
\end{equation}

It is useful to introduce the variable
\begin{equation}
\label{GPPZ_u_def}
  u = 1 - \e{-2r}~,
\end{equation}
in terms of which the following relations hold,
\begin{equation}
\label{GPPZ_W_rels}
\begin{aligned}
  \frac{du}{dr} &= 2(1-u)~, \quad          
  & \e{-2A}     &= \frac{1-u}{u}~,\\
  W             &= -\frac{3}{2u}~,    
  & W_\phi      &= -\sqrt{3} \frac{\sqrt{1-u}}{u}~.
\end{aligned}
\end{equation}

\end{appendix}


\providecommand{\href}[2]{#2}
\begingroup\raggedright
\endgroup

\end{document}